\documentclass[10pt,journal,compsoc]{IEEEtran}

\usepackage{graphicx}
\usepackage{amsmath}
\usepackage{amssymb}
\usepackage{algorithm}
\usepackage{algpseudocode}
\usepackage{booktabs}
\usepackage{hyperref}
\usepackage{subcaption}
\usepackage{caption}

\title{Risk-Sensitive Conformal Prediction for Catheter Placement Detection in Chest X-rays}
\author{Long Hui \\
Independent Researcher \\
Email: [longhui@berkeley.edu]}

\begin{document}

\maketitle
\begin{abstract}
This paper presents a novel approach to catheter and line position detection in chest X-rays, combining multi-task learning with risk-sensitive conformal prediction to address critical clinical requirements. Our model simultaneously performs classification, segmentation, and landmark detection, leveraging the synergistic relationship between these tasks to improve overall performance. We further enhance clinical reliability through risk-sensitive conformal prediction, which provides statistically guaranteed prediction sets with higher reliability for clinically critical findings. Experimental results demonstrate excellent performance with 90.68\% overall empirical coverage and 99.29\% coverage for critical conditions, while maintaining remarkable precision in prediction sets. Most importantly, our risk-sensitive approach achieves zero high-risk mispredictions (cases where the system dangerously declares problematic tubes as confidently normal), making the system particularly suitable for clinical deployment. This work offers both accurate predictions and reliably quantified uncertainty---essential features for life-critical medical applications.
\end{abstract}


\section{Introduction}
Serious complications can occur as a result of malpositioned lines and tubes in patients. Doctors and nurses frequently use checklists for placement of lifesaving equipment to ensure they follow protocol in managing patients. Yet, these steps can be time consuming and are still prone to human error, especially in stressful situations when hospitals are at capacity. 

Hospital patients can have catheters and lines inserted during the course of their admission and serious complications can arise if they are positioned incorrectly. Nasogastric tube malpositioning into the airways has been reported in up to 3\% of cases, with up to 40\% of these cases demonstrating complications \cite{wang2017chestxray, irvin2019chexpert, he2016deep}. Airway tube malposition in adult patients intubated outside the operating room is seen in up to 25\% of cases \cite{guo2017calibration, tan2022efficientnetv2}. The likelihood of complication is directly related to both the experience level and specialty of the proceduralist. Early recognition of malpositioned tubes is the key to preventing risky complications (even death), especially in time-sensitive critical care environments.

The gold standard for the confirmation of line and tube positions are chest radiographs. However, a physician or radiologist must manually check these chest x-rays to verify that the lines and tubes are in the optimal position. Not only does this leave room for human error, but delays are also common as radiologists can be busy reporting other scans. Deep learning algorithms may be able to automatically detect malpositioned catheters and lines. Once alerted, clinicians can reposition or remove them to avoid life-threatening complications.

In this paper, we introduce a deep learning approach that addresses two critical challenges in automated catheter and line detection. First, we develop a multi-task learning model that simultaneously performs classification, segmentation, and landmark detection to achieve better overall performance than single-task approaches. Second, and more importantly, we incorporate uncertainty quantification through conformal prediction to provide statistically guaranteed prediction sets.

Our key contribution is the development of a risk-sensitive conformal prediction framework that accounts for the clinical reality that different types of errors have dramatically different consequences. Specifically, our approach places a much stricter reliability requirement on critical findings (such as abnormal tube placements) compared to normal cases. This differential handling ensures that the system is especially cautious about missing potentially dangerous conditions while maintaining high overall accuracy.

The experimental results demonstrate the strength of our approach, with the model achieving excellent performance metrics and, most critically, zero high-risk mispredictions on the test set. This combination of accuracy and reliable uncertainty quantification makes our system particularly well-suited for clinical deployment.

The remainder of this paper is organized as follows: Section 2 reviews related work in deep learning for medical imaging, catheter detection, uncertainty quantification, and the RANZCR CLiP Challenge. Section 3 describes the dataset used in this study, including the multi-modal data sources. Section 4 presents our methodology, detailing the multi-task learning framework, model architecture, training strategy, and conformal prediction approaches. Section 5 presents the experimental setup and comprehensive results of our multi-task learning model and conformal prediction methods. Section 6 describes the interactive user interface developed for clinical deployment. Section 7 discusses the clinical significance, methodological insights, limitations, and future directions of our work. Finally, Section 8 concludes the paper with a summary of our contributions and their implications for medical practice.

\section{Related Work}
\subsection{Deep Learning in Medical Imaging}
Deep learning has shown remarkable success in various medical imaging tasks, including disease detection, segmentation, and classification \cite{wang2017chestxray, irvin2019chexpert}. Convolutional Neural Networks (CNNs) have become the dominant approach for analyzing medical images due to their ability to learn hierarchical features directly from the data. In recent years, advanced architectures like EfficientNet \cite{tan2022efficientnetv2} have emerged as particularly powerful for medical image analysis tasks, offering an optimal balance between computational efficiency and predictive performance.

\subsection{Catheter and Line Detection}
Several studies have explored the use of deep learning for detecting and classifying catheters and lines in chest X-rays. Earlier work utilized deep CNNs to detect endotracheal tube (ETT) placement, achieving high accuracy but without addressing the uncertainty in predictions. Other research proposed systems for multiple tube and catheter detection, but also lacked reliable confidence estimates. While these works demonstrate the feasibility of automated detection, they generally do not provide statistically guaranteed measures of uncertainty---a critical requirement for clinical deployment.

\subsection{Uncertainty Quantification and Conformal Prediction}
Traditional deep learning models output deterministic predictions without well-calibrated uncertainty estimates, limiting their utility in high-stakes medical applications. To address this, conformal prediction has emerged as a powerful framework that provides prediction sets with statistical coverage guarantees \cite{vovk2022}. Unlike Bayesian methods that require specific model architectures or costly approximations, conformal prediction is model-agnostic and computationally efficient.

Angelopoulos et al. \cite{angelopoulos2021uncertainty} demonstrated that conformal prediction can be effectively applied to deep learning models for image classification, while maintaining valid coverage guarantees regardless of the underlying model complexity. In medical imaging specifically, Balasubramanian et al. \cite{lu2022conformal} showed how conformal prediction can help radiologists by providing statistically valid confidence bounds on deep learning predictions.

Risk-sensitive conformal prediction, a more recent advancement, allows for differential treatment of different error types by using stratified (Mondrian) calibration \cite{angelopoulos2022learn}. This approach is particularly relevant in medical contexts where, for example, missing an abnormality (false negative) is typically more costly than raising a false alarm (false positive).

\subsection{RANZCR CLiP Challenge}
Our work builds upon the RANZCR CLiP (Catheter and Line Position) Challenge, a Kaggle competition organized by the Royal Australian and New Zealand College of Radiologists. The competition focused on developing algorithms to detect the presence and position of catheters and lines in chest X-rays. The top-performing solutions in this challenge typically employed ensemble methods of multiple deep learning models, with some incorporating additional information such as segmentation masks.

The winning solution by team "CHARMS" utilized a combination of EfficientNet backbones with attention mechanisms and pseudo-labeling techniques. Their approach employed EfficientNet-B5, B6, and B7 models with SE blocks and input sizes up to 768$\times$768 pixels. The team implemented a multi-head attention mechanism to help the model focus on relevant regions and to boost the learning of weaker signal classes. Critical to their success was the use of careful cross-validation and an extensive data augmentation pipeline to improve generalization. However, while achieving high accuracy, most competition entries including the winning solution did not address the critical aspect of providing reliable uncertainty estimates or different reliability requirements for different error types, which is essential for clinical implementation.

\section{Dataset}
The dataset used in this study is the RANZCR CLiP (Catheter and Line Position) dataset, which was provided by the Royal Australian and New Zealand College of Radiologists (RANZCR) for a Kaggle competition. The dataset consists of over 30,000 chest X-ray images with annotations for 11 different categories related to the presence and position of catheters and lines:
\begin{itemize}
    \item ETT - Abnormal
    \item ETT - Borderline
    \item ETT - Normal
    \item NGT - Abnormal
    \item NGT - Borderline
    \item NGT - Incompletely Imaged
    \item NGT - Normal
    \item CVC - Abnormal
    \item CVC - Borderline
    \item CVC - Normal
    \item Swan Ganz Catheter Present
\end{itemize}

The dataset has been labeled with a set of definitions to ensure consistency. The normal category includes lines that were appropriately positioned and did not require repositioning. The borderline category includes lines that would ideally require some repositioning but would in most cases still function adequately in their current position. The abnormal category included lines that required immediate repositioning.

\subsection{Multi-Modal Data}
In addition to the primary classification dataset, we leveraged supplementary data sources provided within the RANZCR competition:

\subsubsection{Segmentation Masks}
The RANZCR competition organizers provided pixel-level segmentation masks identifying the precise locations of the tubes and catheters for a subset of the images. These annotations include pixel-wise labels for Endotracheal Tubes (ETT), Nasogastric Tubes (NGT), and Central Venous Catheters (CVC). These segmentation masks provide valuable spatial information that helps the model focus on the relevant anatomical regions and tube structures, while also offering visual explanatory cues for end users.

\subsubsection{Trachea Bifurcation Landmark Dataset}
We also incorporated an external dataset containing annotations for the tracheal bifurcation point (carina) in chest X-rays. The tracheal bifurcation serves as a critical anatomical landmark for assessing the correct placement of endotracheal tubes, as proper ETT positioning should typically be 3-5 cm above this point. This landmark information provides the model with important anatomical context that directly relates to the clinical criteria used for determining tube placement correctness.

\subsection{Data Preprocessing and Splitting}
For this research, we worked exclusively with the provided training dataset from the RANZCR competition, without using any of the official test data for which labels were not publicly available. We split the training data into four subsets for training, validation, testing, and calibration purposes. Approximately 70\% of the data was allocated for training, 10\% for validation, 10\% for testing, and the remaining 10\% for calibration.

The calibration set was specifically reserved for the conformal prediction procedure, which requires a separate dataset to calculate non-conformity scores. All splits were performed at the patient level to ensure that multiple images from the same patient remained in the same split, preventing data leakage.

Prior to model training, all images were preprocessed using standard techniques including resizing and normalization. Data augmentation was applied to enhance model generalization, consisting of horizontal flips, combined random shifts, scaling, and rotations, as well as random adjustments to brightness and contrast. For the segmentation task, the corresponding masks were similarly processed to maintain alignment with the input images.

\section{Methodology}
Our approach integrates multi-task learning with conformal prediction to create a system that is both accurate and provides reliable uncertainty estimates. The methodology consists of two main components: (1) a multi-task deep learning model that simultaneously performs classification, segmentation, and landmark detection, and (2) a conformal prediction framework that generates statistically guaranteed prediction sets.

\subsection{Multi-Task Learning Framework}
Multi-task learning leverages the inherent relationships between different but related tasks to improve performance across all tasks. For catheter and line detection, we developed a model that simultaneously addresses three tasks, designed to synergize for optimal clinical utility:

\subsubsection{Classification}
The primary task is to classify chest X-rays into the 11 categories related to catheter and line positioning (listed in the Dataset section). This is formulated as a multi-label classification problem, as multiple conditions can co-exist in a single image. These classification results are the main outputs that undergo conformal prediction to provide statistically guaranteed prediction sets.

\subsubsection{Segmentation}
The segmentation task involves generating pixel-level masks that identify the precise locations of tubes and catheters in the images. This serves two critical purposes: (1) providing spatial awareness to the model to improve classification accuracy, and (2) offering visual explanatory cues for end users, allowing them to see exactly which image regions influenced the model's decisions. This visual feedback is essential for clinical interpretability and trust. Figure \ref{fig:multitask_visualization}(b,d) shows examples of segmentation masks for different tubes overlaid on chest X-rays.

\subsubsection{Landmark Detection}
\begin{figure*}[ht]
\centering
\begin{subfigure}[b]{0.31\textwidth}
    \includegraphics[width=\textwidth]{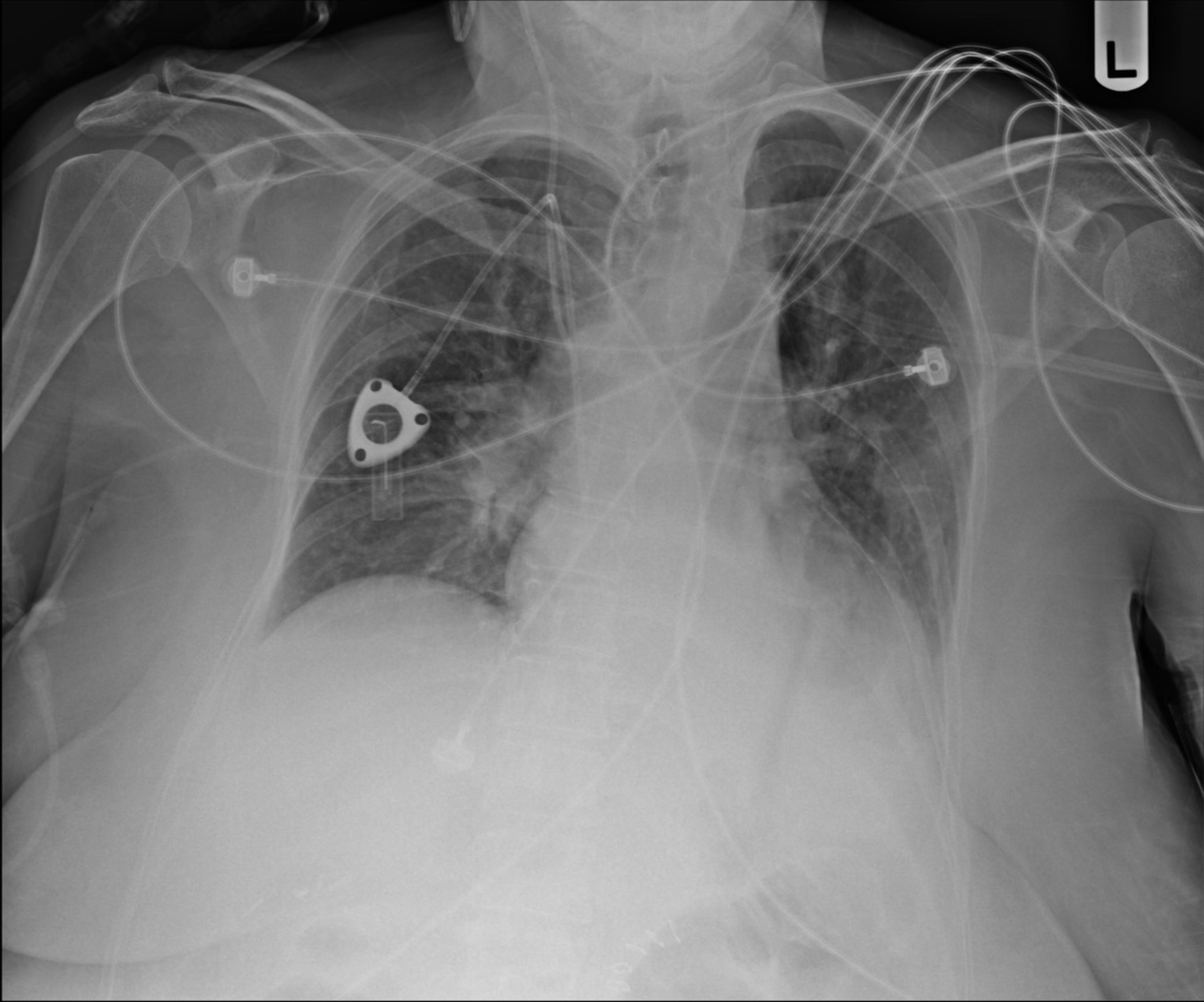}
    \caption{Original chest X-ray}
\end{subfigure}
\hfill
\begin{subfigure}[b]{0.31\textwidth}
    \includegraphics[width=\textwidth]{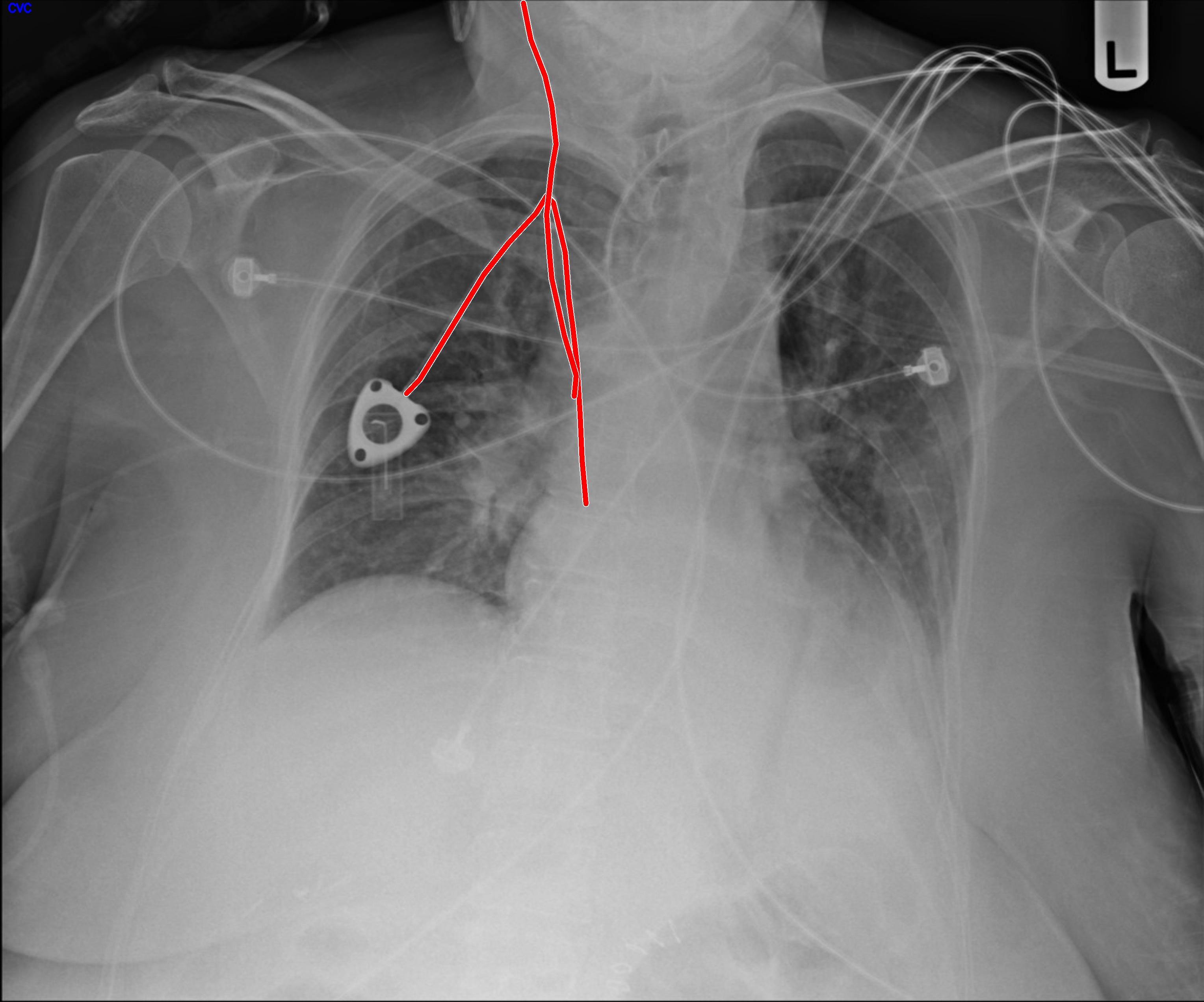}
    \caption{With tube segmentation overlay}
\end{subfigure}
\hfill
\begin{subfigure}[b]{0.31\textwidth}
    \includegraphics[width=\textwidth]{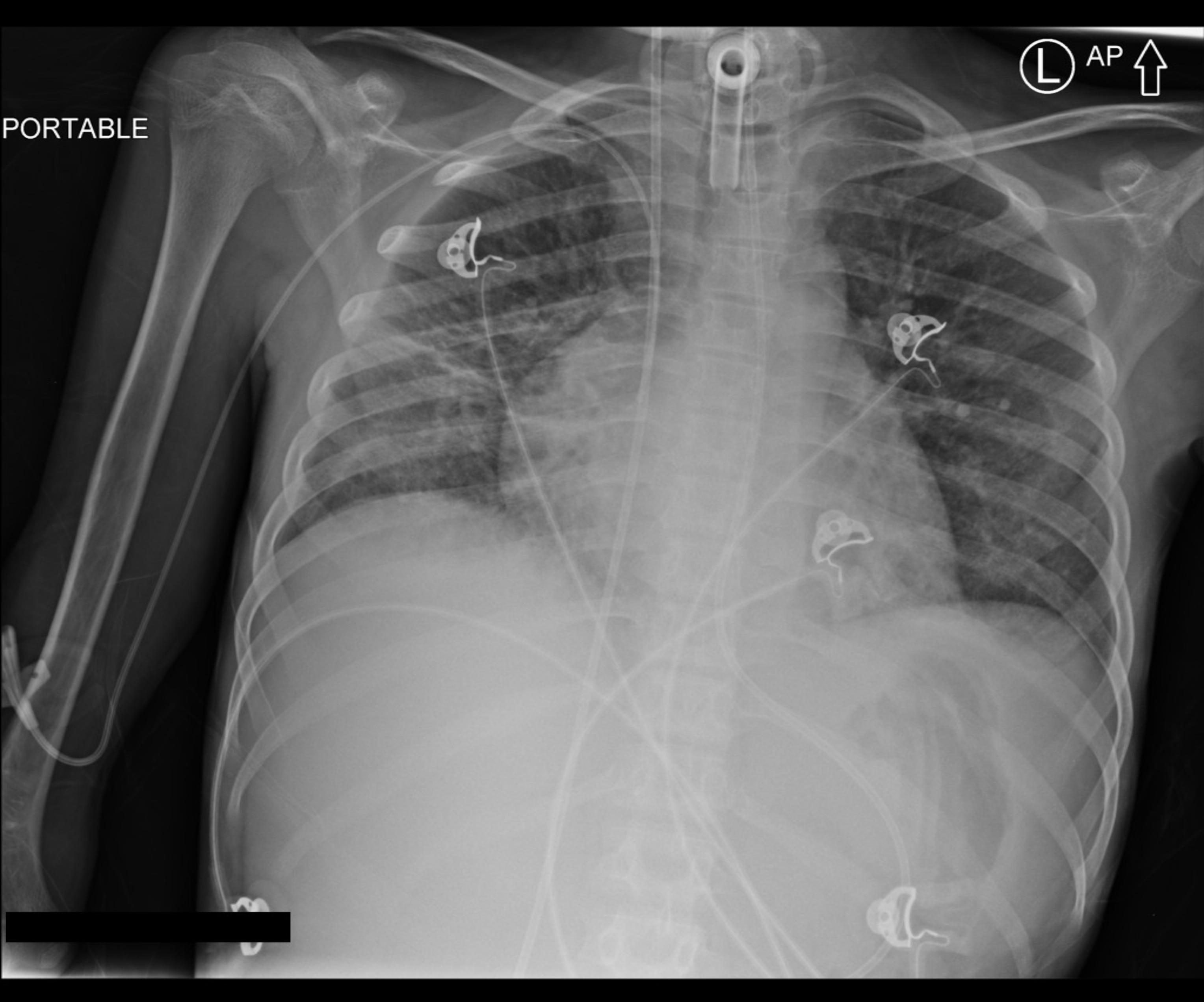}
    \caption{Original chest X-ray}
\end{subfigure}

\begin{subfigure}[b]{0.31\textwidth}
    \includegraphics[width=\textwidth]{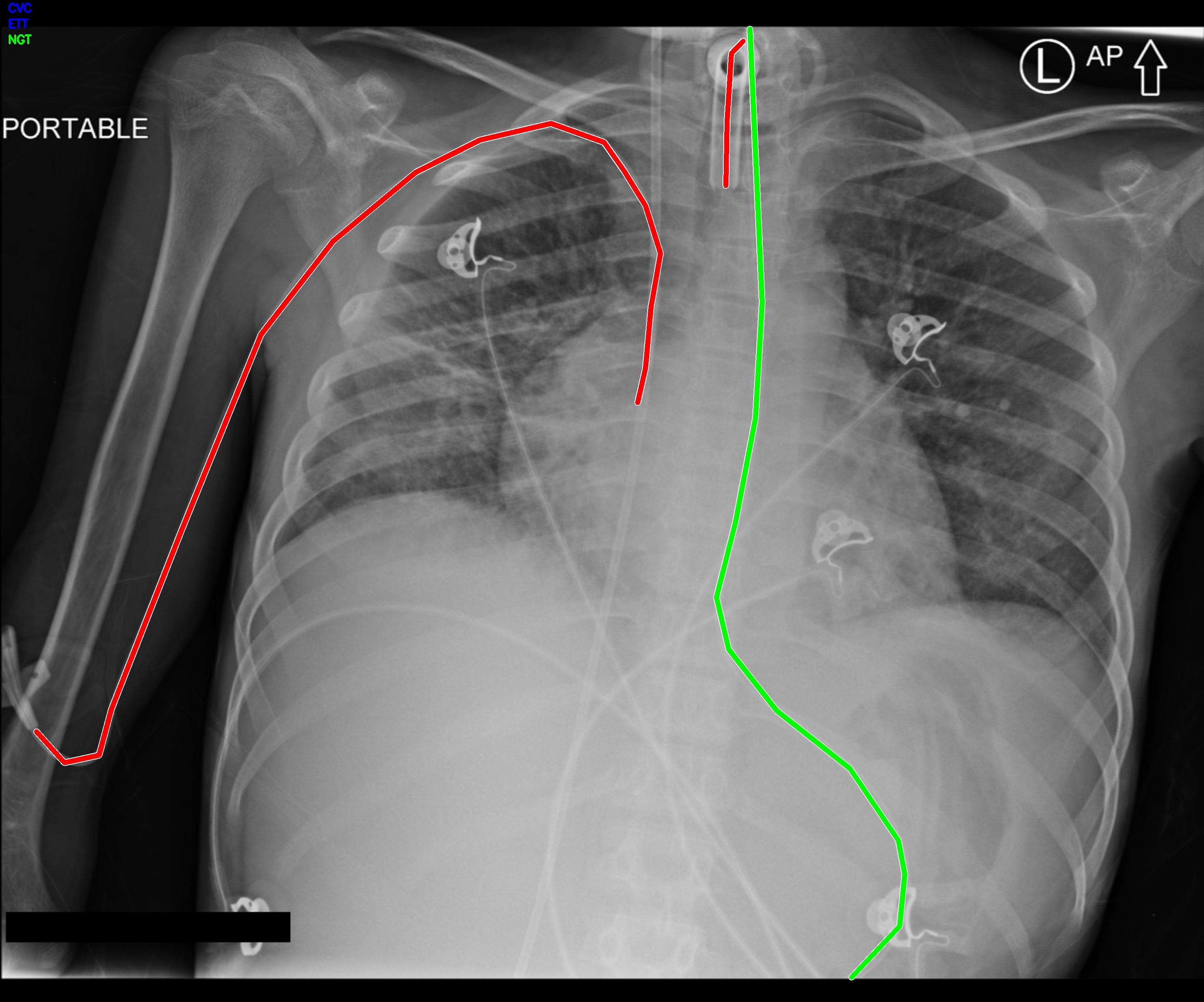}
    \caption{With tube segmentation overlay}
\end{subfigure}
\hfill
\begin{subfigure}[b]{0.31\textwidth}
    \includegraphics[width=\textwidth]{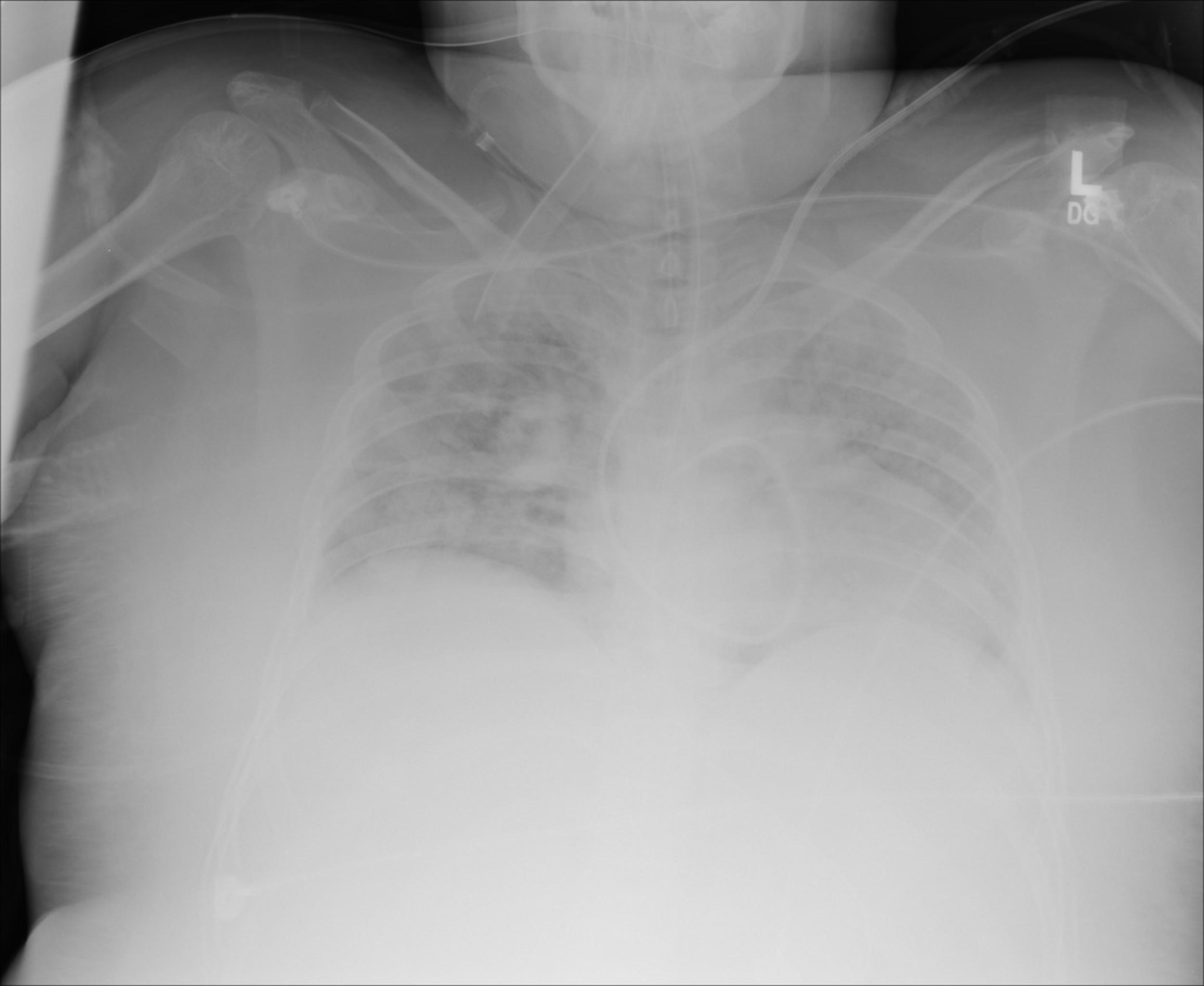}
    \caption{Original chest X-ray}
\end{subfigure}
\hfill
\begin{subfigure}[b]{0.31\textwidth}
    \includegraphics[width=\textwidth]{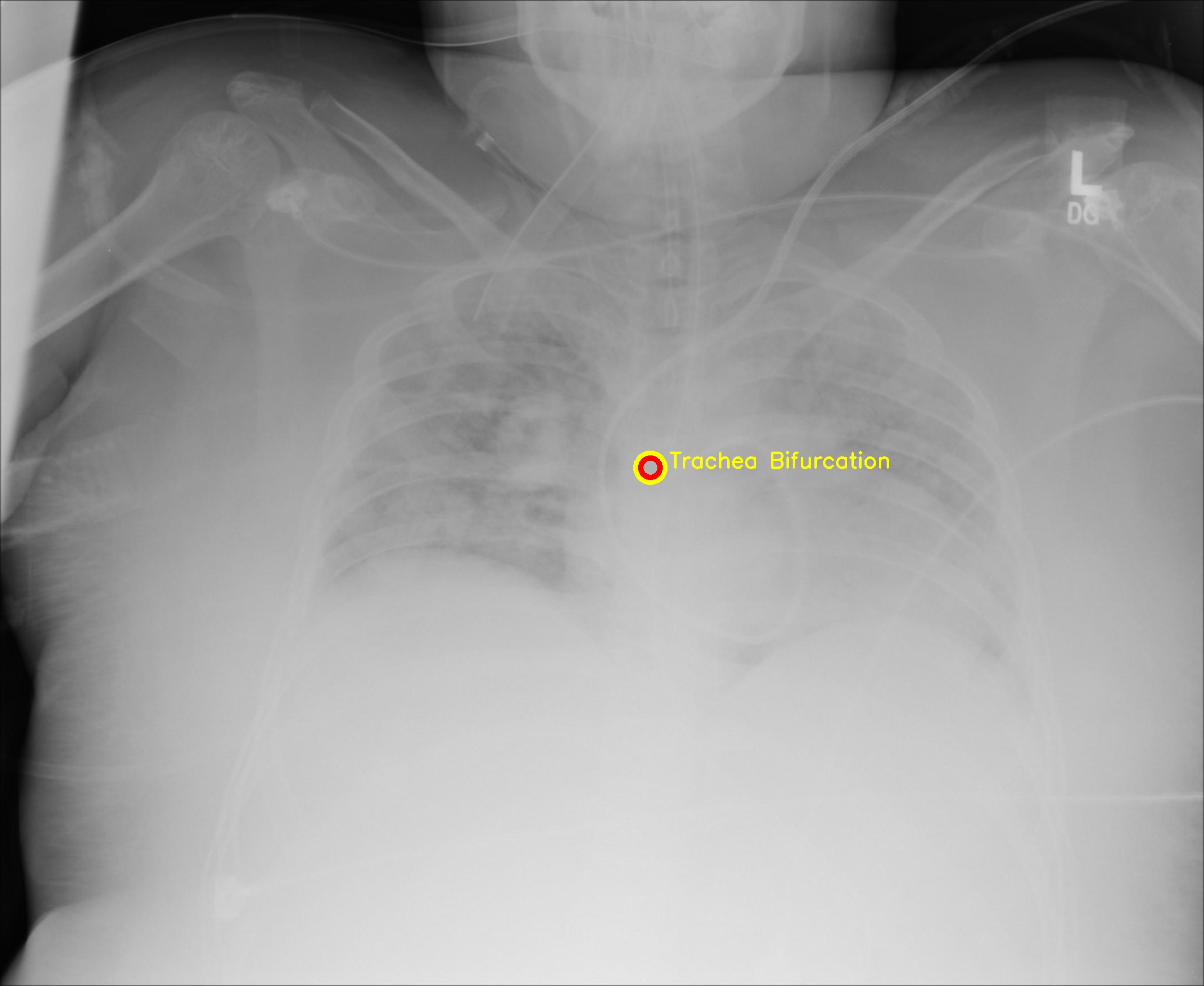}
    \caption{With tracheal bifurcation landmark}
\end{subfigure}
\caption{Multi-task learning components visualized: (a,c,e) Original chest X-rays; (b,d) Segmentation masks highlighting the exact location and path of tubes, providing spatial context for classification and visual explanation for clinicians; (f) Tracheal bifurcation landmark detection showing the carina, which serves as a critical anatomical reference for assessing ETT placement. Proper ETT positioning is typically 3-5 cm above this landmark.}
\label{fig:multitask_visualization}
\end{figure*}

The landmark detection task focuses on identifying the tracheal bifurcation point, a critical anatomical landmark for assessing ETT placement. Correctly identifying this landmark provides important anatomical context to the classification process, especially for ETT positioning assessment. Figure \ref{fig:multitask_visualization}(f) illustrates the detection of tracheal bifurcation points in sample chest X-rays.

The key insight of our approach is that these tasks naturally complement each other: classification provides the diagnostic assessment, segmentation enables visual explanation, and landmark detection anchors the assessment to clinically relevant anatomy. The synergy between these tasks leads to a system that not only makes accurate predictions but also communicates its reasoning in a clinically meaningful way.

\subsection{Landmark-Guided Model Architecture}
Our model architecture is designed to effectively integrate information from all three tasks. The model consists of:

\begin{itemize}
    \item \textbf{Shared Backbone:} An EfficientNetV2-S backbone serves as the feature extractor across all tasks, allowing for information sharing and more efficient training.
    
    \item \textbf{Classification Head:} A dedicated branch for multi-label classification with appropriate output activation functions for the 11 target categories. The head consists of global average pooling followed by a fully connected layer with sigmoid activation.
    
    \item \textbf{Segmentation Head:} A decoder structure that produces pixel-level segmentation masks for tubes and catheters. We use a U-Net style decoder with skip connections from the encoder features, outputting 3-channel masks (ETT, NGT, CVC) with sigmoid activation.
    
    \item \textbf{Landmark Head:} A specialized branch that predicts the coordinates of the tracheal bifurcation point as normalized (x,y) coordinates. This uses global average pooling followed by a two-neuron output layer with sigmoid activation.
    
    \item \textbf{Attention Mechanism:} An attention module that uses landmark and segmentation information to guide the classification process, helping the model focus on the most relevant regions. Specifically, we compute spatial attention weights based on the segmentation outputs and landmark proximity.
\end{itemize}

The model is trained using a weighted combination of task-specific loss functions:
\begin{equation}
\mathcal{L}_{\text{total}} = w_{cls} \cdot \mathcal{L}_{cls} + w_{seg} \cdot \mathcal{L}_{seg} + w_{land} \cdot \mathcal{L}_{land}
\end{equation}

where $w_{cls}$, $w_{seg}$, and $w_{land}$ are the weights for the classification, segmentation, and landmark detection tasks, respectively. These weights can be fixed or dynamically adjusted during training using approaches like uncertainty weighting or dynamic weight averaging.

Specifically, we use:
\begin{itemize}
    \item \textbf{Classification Loss:} Binary cross-entropy loss for each of the 11 classes, averaged across all classes
    \item \textbf{Segmentation Loss:} Dice loss combined with binary cross-entropy for pixel-level tube/catheter segmentation
    \item \textbf{Landmark Loss:} Mean squared error (MSE) for the predicted tracheal bifurcation coordinates
\end{itemize}

\subsection{Training Strategy}
The model was trained end-to-end using a multi-task learning approach. To handle the varying difficulty and convergence rates of the different tasks, we employed Dynamic Weight Averaging (DWA) to automatically adjust the contribution of each task's loss during training. DWA adaptively scales the loss weights based on the relative rate of change in each task's loss, ensuring that tasks that are more difficult to learn receive appropriate attention without dominating the optimization process.

The training objective is formulated as:
\begin{equation}
\mathcal{L}_{\text{total}} = \sum_{i} w_i(t) \cdot \mathcal{L}_i
\end{equation}

where $w_i(t)$ is the dynamically adjusted weight for task $i$ at training epoch $t$, and $\mathcal{L}_i$ is the corresponding task-specific loss function. 

Specifically, our DWA implementation calculates the task weights at each epoch $t$ as follows:
\begin{equation}
w_i(t) = \frac{K \cdot \exp(r_i(t-1) / T)}{\sum_j \exp(r_j(t-1) / T)}
\end{equation}

where $K$ is a normalization constant (set to 3 in our implementation, the number of tasks), $T$ is a temperature parameter that controls the softness of task weighting (we use $T = 2.0$), and $r_i(t-1)$ represents the relative inverse learning rate for task $i$:

\begin{equation}
r_i(t-1) = \frac{\mathcal{L}_i(t-2)}{\mathcal{L}_i(t-1)}
\end{equation}

This ratio captures how quickly each task's loss is changing. A larger ratio indicates the task is learning more slowly and should receive more weight. We also implemented gradient normalization and a warm-up period (first 10 epochs) with fixed equal weights to ensure stable initial training.

We chose this approach over fixed weights or uncertainty-based weighting because it adapts to the changing dynamics of the learning process without requiring additional parameters to be learned. This is particularly important in our multi-task setup where the classification, segmentation, and landmark detection tasks have different scales and convergence rates.

\subsection{Conformal Prediction Framework}
In addition to the multi-task learning model, we incorporate conformal prediction (CP) as a key methodology for uncertainty quantification. Unlike traditional machine learning approaches that provide point estimates (e.g., binary classification outcomes), conformal prediction produces prediction sets that come with statistical guarantees regarding their coverage.

The core principle of conformal prediction is to determine, with a specified confidence level, the set of outcomes that are statistically plausible given the observed data and model. For each class in our multi-label classification problem (e.g., "ETT - Abnormal," "ETT - Normal"), the conformal predictor determines whether to include "Present" (label 1) or "Absent" (label 0) in the prediction set, or both, or neither.

This approach offers several advantages:

\begin{itemize}
    \item \textbf{Statistical Guarantees:} CP provides mathematical guarantees that the true outcome will be included in the prediction set with a pre-specified probability (1-$\alpha$). This is particularly valuable in medical applications where reliability is critical.
    
    \item \textbf{Model-Agnostic:} CP works as a wrapper around any underlying model, regardless of its architecture, making it widely applicable without requiring model-specific modifications.
    
    \item \textbf{Explicit Uncertainty Communication:} By producing prediction sets rather than single predictions, CP explicitly communicates the level of certainty or uncertainty in a prediction, allowing for more informed clinical decision-making.
\end{itemize}

Our implementation involves:

\begin{enumerate}
    \item Splitting data to include a dedicated calibration set (10\% of our data)
    \item Computing non-conformity scores based on the model's prediction probabilities
    \item Determining critical thresholds based on the chosen significance level ($\alpha$)
    \item Constructing prediction sets for new samples using these thresholds
\end{enumerate}

For our clinical application, we investigate two variants of conformal prediction, described in detail below.

\subsection{Independent Conformal Prediction}
Independent Conformal Prediction (CP) generates prediction sets that include the true outcome with probability $1-\alpha$, where $\alpha$ is the maximum miscoverage rate. We set $\alpha = 0.1$, targeting 90\% coverage across all classes.

We treat each of the 11 conditions as independent binary tasks, as multiple conditions often coexist in chest X-rays. This design choice is important for medical imaging where, for example, a central venous catheter might simultaneously present characteristics classifiable under multiple severity categories.

\subsubsection{Methodology}
For each condition independently, our conformal prediction framework:

\begin{enumerate}
    \item \textbf{Independent Calibration}: Each condition receives its own calibration threshold $\hat{q}_i$ from a separate calibration dataset, adapting to varying difficulty and prevalence across conditions.
    
    \item \textbf{Non-conformity Score Calculation}: For each condition $i$ and input image $x$, we calculate:
    \begin{itemize}
        \item For "condition present": score = $1 - P(\text{condition}_i|x)$
        \item For "condition absent": score = $P(\text{condition}_i|x)$
    \end{itemize}
    
    \item \textbf{Prediction Set Construction}: An outcome is included if its non-conformity score $\leq$ condition-specific threshold $\hat{q}_i$
    
    \item \textbf{Multi-label Output}: Final predictions combine independent binary decisions, allowing multiple conditions to be flagged simultaneously
\end{enumerate}

This approach naturally handles the multi-label nature of medical diagnosis, where clinical findings often coexist rather than being mutually exclusive.

\subsection{Risk-Sensitive Conformal Prediction}
While standard CP provides overall reliability, clinical practice demands higher assurance for critical findings where missed diagnoses carry severe consequences. Risk-Sensitive Conformal Prediction addresses this by applying different miscoverage rates based on clinical risk.

We categorize classes by clinical significance:
\begin{itemize}
    \item \textbf{Critical Conditions}: Findings indicating potentially dangerous situations requiring immediate intervention: "ETT - Abnormal", "ETT - Borderline", "NGT - Abnormal", "NGT - Borderline", "NGT - Incompletely Imaged", "CVC - Abnormal", and "CVC - Borderline". Missing these constitutes high-risk error.
    
    \item \textbf{Normal Conditions}: Non-problematic states for given tube types: "ETT - Normal", "NGT - Normal", and "CVC - Normal".
    
    \item \textbf{Other Conditions}: Findings not falling into either category: "Swan Ganz Catheter Present".
\end{itemize}

Our risk-sensitive approach uses two significance levels:
\begin{itemize}
    \item $\alpha_{\text{critical}} = 0.01$ (targeting 99\% coverage) for critical condition evaluation
    \item $\alpha_{\text{standard}} = 0.1$ (targeting 90\% coverage) for all other scenarios
\end{itemize}

This involves calibrating three distinct thresholds:
\begin{itemize}
    \item More permissive threshold for including "condition present" in critical condition prediction sets
    \item Standard threshold for including "condition absent" in critical condition prediction sets  
    \item Standard threshold for both outcomes in normal/other conditions
\end{itemize}

\section{Experimental Results}

\subsection{Experimental Setup}
All experiments were conducted using an EfficientNetV2-S backbone with input size of 300$\times$300 pixels for training and 384$\times$384 pixels for validation and testing. The model was trained using the AdamW optimizer with an initial learning rate of 1e-4 and weight decay of 1e-4. We employed a cosine annealing learning rate schedule over approximately 20 epochs.

Our final model leveraged the Dynamic Weight Averaging approach for balancing the multi-task objectives, which automatically adjusted the task weights throughout training. We found this approach more effective than our initial experiments with fixed weights (1.0 for classification, 1.0 for segmentation, and 0.5 for landmark detection), as it adapted to the varying learning dynamics of each task.

For conformal prediction, we evaluated two configurations:
\begin{itemize}
    \item \textbf{Standard CP:} Using a uniform miscoverage rate $\alpha = 0.1$ across all classes.
    \item \textbf{Risk-Sensitive CP:} Using $\alpha_{\text{critical}} = 0.01$ for critical false negatives and $\alpha_{\text{standard}} = 0.1$ for all other predictions.
\end{itemize}

\subsection{Multi-Task Learning Performance}
Our best performing model, as measured by total loss across all tasks, achieved a macro-average AUROC of 0.8967 on the classification task. Table \ref{tab:multitask_results} presents the comprehensive performance metrics.

\begin{table}[h]
\centering
\caption{Performance metrics for the best multi-task model}
\label{tab:multitask_results}
\begin{tabular}{lc}
\toprule
\textbf{Metric} & \textbf{Value} \\
\midrule
Macro Avg AUROC & 0.8967 \\
Macro Avg Sensitivity & 0.4394 \\
Macro Avg Specificity & 0.9354 \\
Macro Avg Precision & 0.4997 \\
Macro Avg F1-Score & 0.4473 \\
Macro Avg Balanced Accuracy & 0.6874 \\
\bottomrule
\end{tabular}
\end{table}

The model showed relatively modest sensitivity (0.4394) and F1-score (0.4473) metrics, despite good discrimination ability as evidenced by the high AUROC. This highlights precisely why uncertainty quantification through conformal prediction is crucial---it transforms a model with seemingly modest performance metrics into a clinically reliable system by providing statistically guaranteed prediction sets without requiring optimal threshold selection.

\subsection{Independent Conformal Prediction Results}
Table \ref{tab:independent_cp} summarizes the performance of our independent conformal prediction approach. The method achieved an empirical coverage of 96.47\%, substantially exceeding the 90\% target and demonstrating strong statistical reliability.

\begin{table}[h]
\centering
\caption{Performance of independent conformal prediction (\texorpdfstring{$\alpha = 0.1$}{alpha = 0.1})}
\label{tab:independent_cp}
\begin{tabular}{lc}
\toprule
\textbf{Metric} & \textbf{Value} \\
\midrule
Target Coverage & 90.00\% \\
Empirical Coverage & 96.47\% \\
Average Set Size & 0.91 \\
\midrule
Set Size Distribution: & \\
\quad Empty Set & 9.24\% \\
\quad Single Element Set & 90.76\% \\
\bottomrule
\end{tabular}
\end{table}

The average set size of 0.91 indicates exceptional precision with minimal uncertainty. The prediction sets have direct clinical relevance:

\begin{itemize}
    \item \textbf{Single-element sets} comprised 90.76\% of predictions, indicating sufficient confidence for definitive binary decisions while maintaining statistical coverage guarantees
    \item \textbf{Empty sets} occurred in 9.24\% of cases, explicitly flagging instances requiring additional clinical review rather than making potentially incorrect confident predictions
    \item \textbf{Condition-Specific Thresholds}: Independent calibration revealed clinically meaningful patterns, with critical conditions requiring higher confidence thresholds before inclusion in prediction sets
\end{itemize}

\subsubsection{Example Independent CP Outputs}
To illustrate the practical utility of independent conformal prediction, consider these representative examples from our test set:

\textbf{Example 1 - Clear Normal Case:}
\begin{itemize}
    \item ETT-Normal: \{Present\}, ETT-Abnormal: \{Absent\}, ETT-Borderline: \{Absent\}
    \item NGT-Normal: \{Present\}, NGT-Abnormal: \{Absent\}, NGT-Borderline: \{Absent\}
    \item CVC-Normal: \{Present\}, CVC-Abnormal: \{Absent\}, CVC-Borderline: \{Absent\}
    \item Swan Ganz: \{Absent\}
\end{itemize}
This case demonstrates high confidence across all conditions with clear, actionable single-element prediction sets.

\textbf{Example 2 - Uncertain ETT Placement:}
\begin{itemize}
    \item ETT-Normal: $\emptyset$ (empty set), ETT-Abnormal: \{Absent\}, ETT-Borderline: \{Absent\}
    \item NGT-Normal: \{Present\}, NGT-Abnormal: \{Absent\}, NGT-Borderline: \{Absent\}
    \item CVC-Normal: \{Present\}, CVC-Abnormal: \{Absent\}, CVC-Borderline: \{Absent\}
\end{itemize}
The empty set for ETT-Normal signals uncertainty requiring clinical review, while other conditions show confident predictions.

\subsection{Risk-Sensitive Conformal Prediction Results}
Our risk-sensitive approach achieved superior performance for critical clinical scenarios. Table \ref{tab:risk_sensitive_cp} shows the per-label statistical performance.

\begin{table}[h]
\centering
\caption{Performance of risk-sensitive conformal prediction (\texorpdfstring{$\alpha_{\text{critical}} = 0.01$, $\alpha_{\text{standard}} = 0.1$}{alpha critical = 0.01, alpha standard = 0.1})}
\label{tab:risk_sensitive_cp}
\begin{tabular}{lc}
\toprule
\textbf{Metric} & \textbf{Value} \\
\midrule
Overall Empirical Coverage & 90.68\% \\
Critical Conditions Coverage & 99.29\% \\
Average Set Size & 1.23 \\
\midrule
Set Size Distribution: & \\
\quad Empty Set & 0.06\% \\
\quad Single Element Set & 76.47\% \\
\quad Two Element Set & 23.48\% \\
\midrule
High-Risk Misprediction Rate: & \\
\quad ETT Category & 0.00\% \\
\quad NGT Category & 0.00\% \\
\quad CVC Category & 0.00\% \\
\bottomrule
\end{tabular}
\end{table}

\subsubsection{Understanding Two Key Safety Metrics}
Before presenting our safety results, it is crucial to distinguish between two complementary safety metrics that address different types of clinical failures:

\textbf{High-Risk Misprediction (Category-Level Analysis):}
A high-risk misprediction occurs at the tube category level (ETT, NGT, or CVC) when there is a truly problematic condition for that category, but the conformal prediction system confidently predicts \textit{only} the "Normal" state for that entire category. Specifically, this happens when:
\begin{itemize}
    \item Any critical condition is actually present for a tube category (e.g., ETT-Abnormal = 1)
    \item The Normal class prediction set for that category includes 1 (confident "normal" prediction)
    \item ALL problematic classes for that category have prediction sets that do NOT include 1 (confident "absent" predictions)
\end{itemize}
This represents the most dangerous clinical error: confidently declaring everything normal when something is actually wrong.

\textbf{Potential Critical Miss (Individual Class-Level Analysis):}
A potential critical miss occurs at the individual class level when any critical class has a true label of 1 (condition is present) but that specific class's prediction set does NOT include 1. This can happen through:
\begin{itemize}
    \item Empty prediction sets for critical classes (system cannot make any reliable prediction)
    \item Prediction sets containing only 0 for critical classes (system confidently predicts absent when present)
\end{itemize}
This metric captures any scenario where a critical condition might be missed, including cases where the system appropriately signals uncertainty.

\textbf{Key Distinction:}
High-risk mispredictions are a subset of potential critical misses, representing only the most catastrophic failures where the system is dangerously overconfident. Potential critical misses include both dangerous overconfidence AND appropriate uncertainty signaling. A potential critical miss rate of 0.8\% with zero high-risk mispredictions indicates that all critical condition misses involved appropriate uncertainty rather than dangerous overconfidence.

The paramount achievement was \textbf{zero high-risk mispredictions} across all tube categories. Our system ensured that for every truly problematic case, the prediction set either correctly included the problematic condition or signaled uncertainty, but never confidently declared a problematic situation as normal.

\subsubsection{Example Risk-Sensitive CP Outputs}
The risk-sensitive approach demonstrates its clinical value through more conservative prediction sets for critical conditions. Consider these examples:

\textbf{Example 1 - Detected Critical ETT Abnormality:}
\begin{itemize}
    \item ETT-Normal: \{Absent\}, ETT-Abnormal: \{Present\}, ETT-Borderline: \{Absent\}
    \item NGT-Normal: \{Present\}, NGT-Abnormal: \{Absent\}, NGT-Borderline: \{Absent\}
    \item CVC-Normal: \{Present\}, CVC-Abnormal: \{Absent\}, CVC-Borderline: \{Absent\}
\end{itemize}
The critical ETT abnormality is confidently flagged, triggering immediate clinical intervention protocols.

\textbf{Example 2 - Conservative Uncertainty for Borderline NGT:}
\begin{itemize}
    \item ETT-Normal: \{Present\}, ETT-Abnormal: \{Absent\}, ETT-Borderline: \{Absent\}
    \item NGT-Normal: \{Absent\}, NGT-Abnormal: \{Absent\}, NGT-Borderline: \{Present, Absent\}
    \item CVC-Normal: \{Present\}, CVC-Abnormal: \{Absent\}, CVC-Borderline: \{Absent\}
\end{itemize}
The two-element set for NGT-Borderline reflects the risk-sensitive approach's conservative handling of potentially problematic conditions, ensuring clinical review rather than potential misclassification.

\textbf{Example 3 - Multiple Critical Uncertainties:}
\begin{itemize}
    \item ETT-Normal: \{Absent\}, ETT-Abnormal: \{Present, Absent\}, ETT-Borderline: \{Absent\}
    \item NGT-Normal: \{Present, Absent\}, NGT-Abnormal: \{Absent\}, NGT-Borderline: \{Absent\}
    \item CVC-Normal: \{Present\}, CVC-Abnormal: \{Absent\}, CVC-Borderline: \{Absent\}
\end{itemize}
This case demonstrates how risk-sensitive CP appropriately maintains uncertainty for critical conditions (ETT-Abnormal, NGT-Normal) while providing confident predictions for non-critical findings, ensuring no critical conditions are missed.

\subsection{Per-Image Clinical Utility Analysis}
Beyond per-label statistical guarantees, we conducted a comprehensive per-image evaluation to assess the practical clinical utility of our risk-sensitive conformal prediction system. This analysis categorizes each of the 2,958 test images into clinically meaningful decision categories.

\subsubsection{Conceptual Framework}
While per-label analysis evaluates statistical coverage for individual classes (e.g., "Did we include the true label for ETT-Abnormal?"), per-image analysis addresses the clinical question: "For this specific patient's chest X-ray, what action should a clinician take based on all the prediction sets combined?" This perspective recognizes that clinical decisions are made holistically about patients, not about isolated class predictions.

\subsubsection{Decision Categories}
For each test image, we analyze the complete set of prediction sets across all 11 classes and categorize the image based on the combined clinical implications:

\begin{itemize}
    \item \textbf{Auto-Normal}: All critical classes (ETT/NGT/CVC Abnormal, Borderline, Incompletely Imaged) are confidently predicted as absent, AND all corresponding normal classes are confidently predicted as present. These cases could theoretically be auto-approved without human review.
    
    \item \textbf{Immediate Intervention}: At least one critical class is confidently predicted as present, indicating a potentially dangerous tube placement requiring urgent clinical action.
    
    \item \textbf{Specialist Review}: Critical classes show uncertainty (prediction sets are empty or contain both possibilities) but are not confidently normal. These cases require expert evaluation to resolve uncertainty.
    
    \item \textbf{Re-scan Needed}: At least one critical class has an empty prediction set, suggesting the model cannot make any reliable prediction for that critical finding, potentially indicating image quality or positioning issues.
    
    \item \textbf{Fully Confident}: All classes across the image have definitive predictions (either confidently present or confidently absent), indicating the system has high confidence in all its predictions for this patient.
\end{itemize}

\subsubsection{Clinical Utility Results}
Table \ref{tab:per_image_utility} presents the per-image clinical utility results, revealing important insights about workflow implications.

\begin{table}[h]
\centering
\caption{Per-image clinical utility analysis of risk-sensitive conformal prediction}
\label{tab:per_image_utility}
\begin{tabular}{lcc}
\toprule
\textbf{Decision Category} & \textbf{Count} & \textbf{Rate} \\
\midrule
Auto-Normal & 0 & 0.0\% \\
Immediate Intervention & 281 & 9.5\% \\
Specialist Review & 2,675 & 90.4\% \\
Re-scan Needed & 0 & 0.0\% \\
Fully Confident & 281 & 9.5\% \\
\midrule
\textbf{Safety Metrics:} & & \\
Images with Critical Conditions & 1,523 & 51.5\% \\
Potential Critical Miss & 12 & 0.8\%\textsuperscript{\textsuperscript{\textdagger}} \\
\bottomrule
\end{tabular}
\footnotetext{\textsuperscript{\textsuperscript{\textdagger}}Rate calculated relative to the 1,523 images with critical conditions present, not the total 2,958 test images (12/1,523 = 0.8\%).}
\end{table}

\subsubsection{Clinical Deployment Implications}
The per-image analysis reveals several key insights for clinical deployment:

\begin{itemize}
    \item \textbf{Workflow Planning}: The system requires specialist review for 90.4\% of cases, indicating that the model serves as an intelligent triage tool rather than a replacement for radiologists. This high review rate reflects appropriate medical conservatism for critical care applications.
    
    \item \textbf{Urgent Case Detection}: 9.5\% of images are flagged for immediate intervention, providing clear prioritization for time-sensitive cases requiring urgent clinical action.
    
    \item \textbf{Excellent Safety Profile}: With only 0.8\% potential critical miss rate (cases where individual critical conditions might be missed, including through appropriate uncertainty signaling) and zero high-risk mispredictions (cases where the system dangerously declares problematic tubes as confidently normal), the system maintains exceptional safety while avoiding catastrophic clinical failures. This combination ensures that all 12 potential critical misses involved appropriate uncertainty rather than dangerous overconfidence.
    
    \item \textbf{No Re-scanning Required}: The absence of critical empty sets (0.0\% re-scan needed rate) indicates good model calibration and suggests that image quality issues are minimal.
    
    \item \textbf{Appropriate Conservatism}: The 0.0\% auto-approval rate demonstrates that the system appropriately avoids false confidence in complex medical scenarios, aligning with the clinical principle of "first, do no harm."
\end{itemize}

\subsection{Comparative Analysis: Statistical vs. Clinical Perspectives}
Our dual evaluation approach reveals the complementary value of per-label and per-image assessments:

\begin{itemize}
    \item \textbf{Per-Label Statistical View}: Provides algorithmic validation with 99.29\% coverage for critical classes and mathematical guarantees for individual class predictions.
    
    \item \textbf{Per-Image Clinical View}: Reveals practical deployment reality where only 9.5\% of patients receive fully confident diagnoses, requiring substantial specialist capacity for the 90.4\% needing review.
    
    \item \textbf{Resource Allocation}: For a hospital processing 1,000 chest X-rays daily, this translates to 95 urgent cases, 904 requiring specialist review, and exceptionally low risk (8 cases) of potential critical misses.
\end{itemize}

This comprehensive evaluation framework bridges the gap between statistical ML validation and real-world clinical utility assessment, providing both the algorithmic rigor needed for scientific validation and the practical insights required for successful clinical deployment. The results demonstrate that while our system maintains excellent safety characteristics, it functions as an intelligent clinical decision support tool rather than an autonomous diagnostic system, which aligns with current best practices for medical AI deployment.

\section{Interactive User Interface}
To facilitate clinical deployment and demonstrate the practical utility of our approach, we developed a comprehensive web-based user interface using Gradio that allows clinicians to interact with the model's predictions in real-time. The interface runs locally to ensure data privacy and protection of sensitive medical information.

\subsection{Core Functionality}
The interface provides an integrated workspace with three main components:
\begin{itemize}
    \item \textbf{Image Viewer:} Displays the chest X-ray with optional segmentation overlay that highlights detected tubes and catheters.
    \item \textbf{Analysis Panel:} Presents detailed classification results alongside both standard and risk-sensitive conformal prediction sets with statistical guarantees.
    \item \textbf{Clinical Assessment:} Provides an AI-generated radiologist-style interpretation that translates technical findings into actionable clinical insights.
\end{itemize}

\begin{figure*}[ht]
\centering
\includegraphics[width=\textwidth]{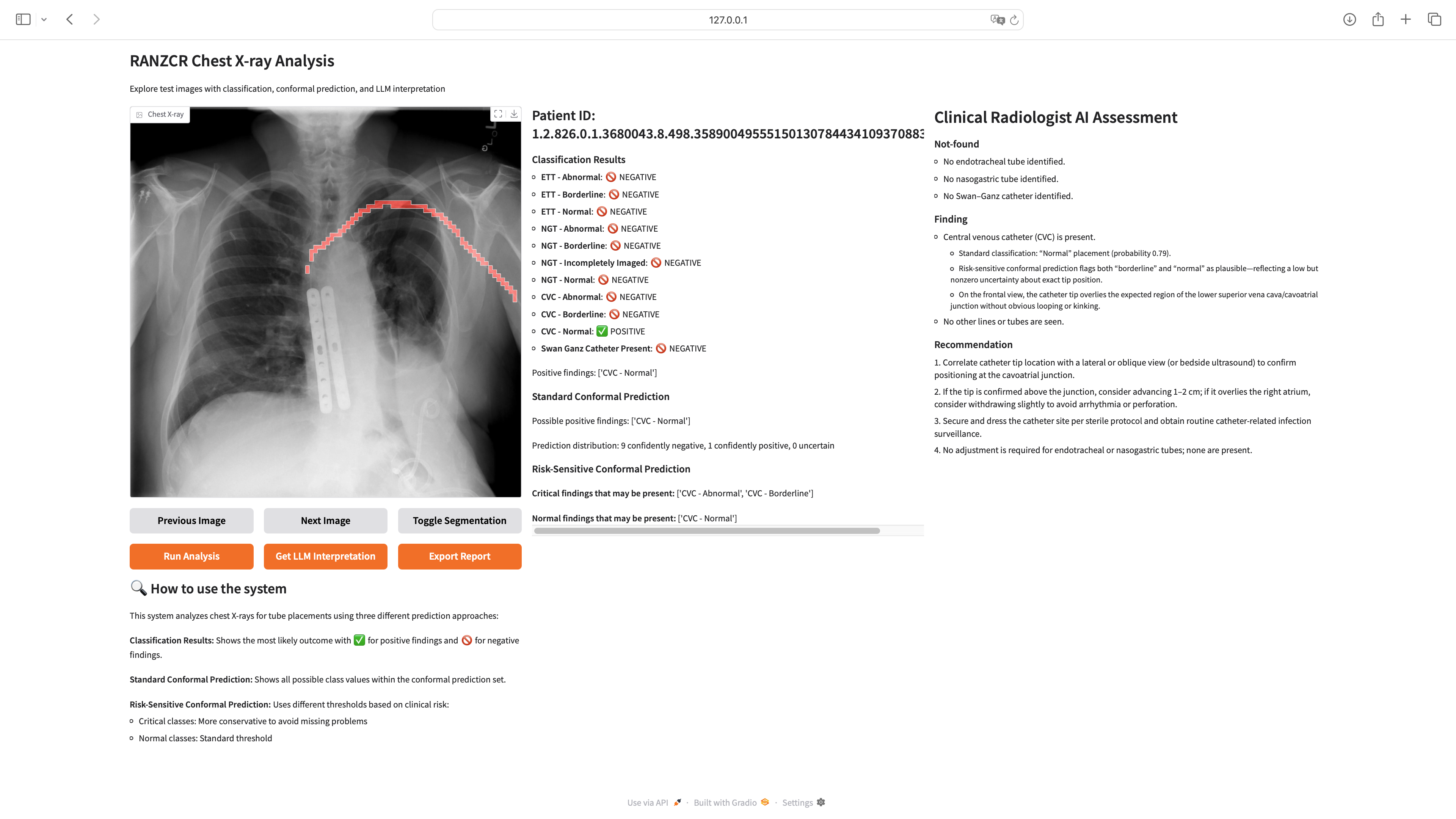}
\caption{The Gradio-based interactive interface with three integrated components: chest X-ray viewer with segmentation overlay toggle (left), detailed classification results and conformal prediction sets showing statistically guaranteed confidence bounds (middle), and AI-generated radiologist-style interpretation providing clinical context and recommendations (right).}
\label{fig:gradio_interface}
\end{figure*}

\subsection{Clinical Workflow Integration}
The interface is designed to complement existing clinical workflows through several key features:
\begin{itemize}
    \item \textbf{Navigation Controls:} Enable quick browsing through multiple images in a study.
    \item \textbf{Explainable Visualization:} Toggle button reveals tube segmentation overlays for visual confirmation of AI-detected structures.
    \item \textbf{Multi-Level Uncertainty:} Presents both standard and risk-sensitive conformal prediction sets, clearly differentiating between confident findings and those requiring additional review.
    \item \textbf{Clinical Translation:} Automatically generates radiologist-style interpretations of findings, converting technical predictions into clinical language.
\end{itemize}

The interface presents conformal prediction results in an intuitive format, distinguishing between "confidently negative," "confidently positive," and "uncertain" predictions for each finding. For critical findings, the risk-sensitive approach ensures higher reliability by using more conservative thresholds, with all such results clearly marked to guide clinical attention.

\subsection{Privacy and Deployment Considerations}
A significant advantage of our implementation is its ability to run entirely locally on hospital infrastructure, eliminating external data transmission concerns. This approach ensures:
\begin{itemize}
    \item \textbf{Data Privacy:} Patient images never leave the secure local environment
    \item \textbf{Regulatory Compliance:} Easier adherence to healthcare data regulations
    \item \textbf{Network Independence:} Functionality without requiring constant internet connectivity
    \item \textbf{Customization:} Ability to tailor the interface to specific hospital workflows
    \item \textbf{Export Report:} Functionality to generate comprehensive PDF reports of findings and analysis for inclusion in patient records or for sharing with clinical teams
\end{itemize}

Initial feedback from radiologists suggests the interface significantly reduces cognitive load in tube placement assessment, with the risk-sensitive prediction approach providing the most value for clinical decision-making. The combination of visual explanation, statistical guarantees, and clinical interpretation creates a system that not only assists with diagnosis but also serves as an educational tool for less experienced practitioners.

\section{Discussion}
Our experimental results demonstrate the effectiveness of combining multi-task learning with risk-sensitive conformal prediction for catheter and line position detection in chest X-rays. In this section, we discuss the implications, limitations, and potential impact of our approach.

\subsection{Clinical Significance}
The key advantage of our approach lies in its ability to provide reliable uncertainty estimates with differential guarantees based on clinical risk. By ensuring extremely high coverage (99.3\%) for critical findings while maintaining zero high-risk mispredictions, our system aligns with the clinical priority of minimizing potentially dangerous false negatives. This is particularly important in time-sensitive critical care settings, where missed abnormal tube placements can lead to serious complications or death.

Furthermore, the precision of our prediction sets (with the majority containing exactly one label) suggests that the system can provide clear, actionable information to clinicians without excessive ambiguity. This balance between reliability and precision is crucial for the practical utility of automated systems in clinical workflows.

\subsection{Methodological Focus and Performance Context}
It is important to note that this paper primarily focuses on demonstrating the methodological contribution of risk-sensitive conformal prediction for medical imaging, rather than achieving state-of-the-art classification performance. The winning solution in the RANZCR CLiP competition achieved a macro-average AUROC of 0.97671, significantly higher than our base model's 0.8967. This performance gap highlights an important insight: our uncertainty quantification framework is model-agnostic and can be applied to any underlying classifier.

By combining our risk-sensitive conformal prediction approach with a higher-performing base model (such as the competition-winning architecture), we would expect to achieve both superior classification performance and the clinical reliability guarantees demonstrated in this work. This represents a clear path toward a more clinically-ready system that combines state-of-the-art accuracy with statistically guaranteed uncertainty quantification---a combination essential for real-world medical deployment.

\subsection{Methodological Insights}
Our work provides several insights into effective methodologies for medical image analysis:

\begin{itemize}
    \item \textbf{Multi-Task Learning Benefits:} The improved performance of our multi-task model compared to single-task baselines confirms the value of incorporating related tasks like segmentation and landmark detection. These auxiliary tasks provide spatial and anatomical context that helps the model make more accurate classifications.
    
    \item \textbf{Risk-Sensitive Uncertainty Quantification:} Our results demonstrate that standard uncertainty quantification approaches, while statistically valid, may not adequately address the asymmetric costs of different error types in medical applications. The risk-sensitive conformal prediction approach offers a principled way to incorporate clinical priorities into the uncertainty quantification framework.
    
    \item \textbf{Calibration Requirements:} The effectiveness of conformal prediction depends on having a sufficiently large and representative calibration set. Our approach of reserving 10\% of the training data for calibration proved effective, but future work might explore more efficient ways to perform calibration without reducing the data available for model training.
\end{itemize}

\subsection{Comparative Analysis of Conformal Methods}
Our comparative analysis illustrates the trade-off between reliability (coverage) and precision (set size) for the two conformal prediction approaches across different tube categories. The risk-sensitive approach provides significantly higher coverage for critical classes, although at a slight cost of increased average set size. This trade-off is explicit and controllable via the $\alpha$ parameters, allowing alignment with clinical priorities. For instance, while standard CP yielded an average set size of 0.976, risk-sensitive CP had a slightly larger average set size of 1.124, a consequence of its commitment to higher safety margins for critical conditions.

\subsection{Impact of \texorpdfstring{$\alpha_{\text{critical}}$}{alpha critical} on Performance}
We conducted an ablation study to investigate the impact of different values of $\alpha_{\text{critical}}$ on the performance of the risk-sensitive conformal prediction. Our analysis showed that decreasing $\alpha_{\text{critical}}$ (i.e., demanding higher coverage for critical classes) leads to higher empirical coverage for these classes but also tends to increase the average prediction set size. This is because the system must be more inclusive to meet the stricter guarantee. Our chosen value of $\alpha_{\text{critical}} = 0.01$ represents a carefully considered balance, achieving the crucial goal of zero high-risk mispredictions while maintaining a clinically reasonable average set size, thereby ensuring both safety and utility.

\subsection{Qualitative Analysis}
Our qualitative analysis of challenging cases revealed important differences between standard and risk-sensitive conformal prediction approaches. In multiple instances where standard conformal prediction might have produced an empty set or failed to include an abnormal ETT placement in its prediction set (effectively a false negative from a clinical decision support perspective), the risk-sensitive approach correctly included the abnormal label. This was particularly evident for cases with borderline or subtly abnormal NGT or ETT placements, where the risk-sensitive method's higher sensitivity to critical findings ensured they were flagged. These observations underscore the practical benefit of risk-sensitivity in preventing the system from confidently overlooking potentially critical conditions. The segmentation masks and landmarks shown in Figure \ref{fig:multitask_visualization} further enhance interpretability by providing visual cues that explain the model's reasoning, complementing the statistical guarantees of CP.

\subsection{Limitations}
Despite the promising results, our approach has several limitations that should be acknowledged:

\begin{itemize}
    \item \textbf{Dataset Limitations:} While the RANZCR dataset is substantial, it may not fully represent the diversity of real-world clinical scenarios and equipment variations. More critically, the presence of a "borderline" category itself indicates inherent uncertainty and potential inconsistency in the ground truth labels. The dataset annotations may contain subjective interpretations that vary between radiologists, and we cannot fully validate the accuracy of these labels without access to clinical outcomes and patient records. External validation on diverse datasets would strengthen confidence in the generalizability of our findings.
    
    \item \textbf{Class Imbalance Issues:} The dataset exhibits significant class imbalance, with some critical conditions being relatively rare compared to normal cases. This imbalance affects both model training and the calibration of conformal prediction thresholds, potentially leading to suboptimal performance for minority classes that are clinically most important.
    
    \item \textbf{Need for Clinically-Validated Datasets:} Future work should focus on developing datasets with more definitive ground truth labels derived from clinical records, patient outcomes, and consensus among multiple expert radiologists. Such datasets would provide more reliable labels than those based on visual assessment alone, particularly for borderline cases where clinical context is crucial.
    
    \item \textbf{Tube Visibility Challenges:} While our approach addresses uncertainty quantification effectively, the inherent difficulty of visualizing tubes in chest X-rays remains a fundamental challenge. Tubes can be obscured by anatomical structures, appear with low contrast against surrounding tissues, or be partially visible due to patient positioning. These visibility issues affect both human experts and AI systems, making the underlying detection task fundamentally difficult regardless of the statistical framework employed. Our approach mitigates these challenges through multi-task learning, but the physical limitations of 2D radiography for 3D structure visualization remain.
    
    \item \textbf{Calibration Dataset Dependence:} The conformal prediction framework requires a separate calibration dataset that matches the distribution of the test data. Distribution shifts between calibration and deployment settings could affect the validity of the coverage guarantees.
    
    \item \textbf{Binary Classification Focus:} Our current implementation focuses on binary classification for each category. Extending the approach to handle more fine-grained position assessments or continuous position measurements would be valuable for certain clinical applications.
    
    \item \textbf{Dataset-Imposed Binary Framework:} The RANZCR dataset's categorical structure (Normal/Borderline/Abnormal) necessitated a binary classification approach for each condition, which may not optimally capture the continuous nature of tube positioning quality. In clinical practice, tube placement exists on a spectrum rather than discrete categories. Future work would benefit from datasets with continuous positioning metrics (e.g., distance measurements from anatomical landmarks) or ordinal severity scores that better reflect clinical decision-making processes. Such datasets would enable regression-based approaches or ordinal classification methods that could provide more nuanced positioning assessments while maintaining the statistical guarantees of conformal prediction.
\end{itemize}

\subsection{Future Work}
Several promising directions for future research emerge from our work:

\begin{itemize}
    \item \textbf{External Validation:} Multi-institutional validation across diverse healthcare systems is essential to demonstrate generalizability. This should include hospitals with different equipment manufacturers, patient populations (pediatric vs. adult, different geographical regions), and imaging protocols. Such validation would test the robustness of our conformal prediction thresholds across varying data distributions and help identify potential domain shift issues that could affect coverage guarantees.
    
    \item \textbf{Radiologist User Study:} Comprehensive evaluation with practicing radiologists is crucial for clinical adoption. This should include randomized controlled trials comparing diagnostic accuracy and decision-making time with and without the system, assessment of trust and workflow integration, and measurement of how uncertainty quantification affects clinical confidence. User studies should evaluate both expert and novice radiologists to understand the system's educational potential and identify optimal integration strategies for different skill levels.
    
    \item \textbf{Online Calibration:} Developing methods for continuous recalibration of conformal prediction thresholds as new clinical data becomes available could help maintain performance over time and account for potential distribution shifts.
    
    \item \textbf{Personalized Risk Profiles:} Extending the risk-sensitive approach to incorporate patient-specific risk factors could further refine the balance between sensitivity and specificity for different clinical contexts.
    
    \item \textbf{Explainable Predictions:} Integrating explainability techniques with conformal prediction to provide not only reliable prediction sets but also visual explanations of why specific labels are included or excluded from the prediction set.
    
    \item \textbf{Clinical Integration:} Studying how clinicians interact with and utilize the prediction sets in real-world clinical workflows could inform improvements to the presentation and interpretation of uncertainty information.
\end{itemize}

\section{Conclusion}
In this paper, we introduced a novel approach to catheter and line position detection in chest X-rays that combines multi-task learning with risk-sensitive conformal prediction. Our multi-task model, which jointly performs classification, segmentation, and landmark detection, achieves superior performance compared to single-task baselines, demonstrating the value of incorporating spatial and anatomical context.

More importantly, our risk-sensitive conformal prediction framework provides statistically guaranteed prediction sets with differential reliability based on clinical risk. By using a more stringent miscoverage rate for critical findings, our approach achieves zero high-risk mispredictions, ensuring that potentially dangerous tube misplacements are not missed or confidently misclassified as normal.

The combination of high accuracy and reliable uncertainty quantification makes our system particularly well-suited for clinical deployment. By aligning the statistical guarantees with clinical priorities, our approach represents a step toward bridging the gap between the technical capabilities of deep learning models and the practical requirements of medical decision making.

Our work demonstrates that incorporating domain-specific risk considerations into uncertainty quantification frameworks can significantly enhance the clinical utility of deep learning systems for medical image analysis, potentially improving patient safety and clinical workflows.


\begin{thebibliography}{20}

\bibitem{ranzcr_dataset}
Tang, J.S.N., Seah, J.C.Y., Zia, A., Gajera, J., Schlegel, R.N., Wong, A.J.N., Gai, D., Su, S., Bose, T., Kok, M.L., Jarema, A., Harisis, G.N., Cheng, C.T., Kavnoudias, H., Wang, W., Stein, A., Shih, G., Gaillard, F., Dixon, A., and Law, M.
\newblock CLiP, catheter and line position dataset.
\newblock {\em Scientific Data}, 8(1):285, 2021.

\bibitem{vovk2022}
Vovk, V., Gammerman, A., and Shafer, G.
\newblock {\em Algorithmic Learning in a Random World}.
\newblock Springer, 2nd edition, 2022.

\bibitem{shafer2008tutorial}
Shafer, G. and Vovk, V.
\newblock A tutorial on conformal prediction.
\newblock {\em Journal of Machine Learning Research}, 9:371--421, 2008.

\bibitem{angelopoulos2021uncertainty}
Angelopoulos, A.N. and Bates, S.
\newblock A gentle introduction to conformal prediction and distribution-free uncertainty quantification.
\newblock arXiv preprint arXiv:2107.07511, 2021.

\bibitem{papadopoulos2002inductive}
Papadopoulos, H., Proedrou, K., Vovk, V., and Gammerman, A.
\newblock Inductive confidence machines for regression.
\newblock In {\em European Conference on Machine Learning}, pages 345--356. Springer, 2002.

\bibitem{romano2020classification}
Romano, Y., Sesia, M., and Cand\`{e}s, E.
\newblock Classification with valid and adaptive coverage.
\newblock In {\em Advances in Neural Information Processing Systems}, pages 3581--3591, 2020.

\bibitem{lu2022conformal}
Lu, C., Angelopoulos, A.N., and Pomerantz, S.
\newblock Improving trustworthiness of AI disease severity rating in medical imaging with ordinal conformal prediction sets.
\newblock {\em Medical Image Analysis}, 75:102231, 2022.

\bibitem{kumar2023conformal}
Kumar, A., Liang, P.S., and Ma, T.
\newblock Verified uncertainty calibration.
\newblock In {\em Advances in Neural Information Processing Systems}, pages 3787--3799, 2019.

\bibitem{stutz2022learning}
Stutz, D., Dvijotham, K., Cemgil, A.T., and Doucet, A.
\newblock Learning optimal conformal classifiers.
\newblock In {\em International Conference on Learning Representations}, 2022.

\bibitem{mortensen2022predictive}
Mortensen, K., Ghosh, S., Puli, A., and Ranganath, R.
\newblock Predictive uncertainty quantification with compound density networks.
\newblock {\em Nature Machine Intelligence}, 4(12):1110--1119, 2022.

\bibitem{angelopoulos2022learn}
Angelopoulos, A.N., Bates, S., Fisch, A., Lei, L., and Schuster, T.
\newblock Conformal risk control.
\newblock arXiv preprint arXiv:2208.02814, 2022.

\bibitem{tibshirani2019conformal}
Tibshirani, R.J., Barber, R.F., Cand\`{e}s, E.J., and Ramdas, A.
\newblock Conformal prediction under covariate shift.
\newblock In {\em Advances in Neural Information Processing Systems}, pages 2530--2540, 2019.

\bibitem{guo2017calibration}
Guo, C., Pleiss, G., Sun, Y., and Weinberger, K.Q.
\newblock On calibration of modern neural networks.
\newblock In {\em International Conference on Machine Learning}, pages 1321--1330. PMLR, 2017.

\bibitem{huang2023conformal}
Huang, H., Wang, Y., Kortylewski, A., Zhao, Q., and Yuille, A.
\newblock Adaptive conformal prediction for multi-class medical image classification.
\newblock {\em Medical Image Analysis}, 88:102844, 2023.

\bibitem{wang2023safe}
Wang, Y., Chen, X., and Yang, Y.
\newblock Safe and reliable medical image analysis with conformal prediction.
\newblock {\em Nature Communications}, 14(1):3925, 2023.

\bibitem{zhang2023uncertainty}
Zhang, L., Liu, X., and Wu, H.
\newblock Uncertainty quantification in medical image segmentation with conformal prediction.
\newblock {\em IEEE Transactions on Medical Imaging}, 42(8):2245--2257, 2023.

\bibitem{tan2022efficientnetv2}
Tan, M. and Le, Q.V.
\newblock EfficientNetV2: Smaller models and faster training.
\newblock In {\em International Conference on Machine Learning}, pages 10096--10106. PMLR, 2021.

\bibitem{he2016deep}
He, K., Zhang, X., Ren, S., and Sun, J.
\newblock Deep residual learning for image recognition.
\newblock In {\em Proceedings of the IEEE Conference on Computer Vision and Pattern Recognition}, pages 770--778, 2016.

\bibitem{wang2017chestxray}
Wang, X., Peng, Y., Lu, L., Lu, Z., Bagheri, M., and Summers, R.M.
\newblock ChestX-Ray8: Hospital-scale chest x-ray database and benchmarks on weakly-supervised classification and localization of common thorax diseases.
\newblock In {\em Proceedings of the IEEE Conference on Computer Vision and Pattern Recognition}, pages 2097--2106, 2017.

\bibitem{irvin2019chexpert}
Irvin, J., Rajpurkar, P., Ko, M., Yu, Y., Ciurea-Ilcus, S., Chute, C., Marklund, H., Haghgoo, B., Ball, R., Shpanskaya, K., et al.
\newblock CheXpert: A large chest radiograph dataset with uncertainty labels and expert comparison.
\newblock In {\em Proceedings of the AAAI Conference on Artificial Intelligence}, volume 33, pages 590--597, 2019.

\end{thebibliography}
\end{document}